\documentstyle[12pt]{article}
\newcommand{\ti}{\tilde}

\newcommand{\ve}{\varepsilon}

\def\s{\sigma}
\def\o{\omega}

\def\ve{\varepsilon}

\def\bba{\begin{array}}
\def\eea{\end{array}}

\def\ra{\rightarrow}
\def\ti{\tilde}

\def\bb{\begin{equation}}
\def\ee{\end{equation}}

\begin{document}\large


\centerline{\LARGE Initial boundary value problem on a half line }
\centerline{\LARGE for the MKdV equation \footnote{This work has been supported
by Russian Foundation for Fundamental Researches, Grant \# 98-01-00576}}

\bigskip
\bigskip
\centerline{I.T.Habibullin}
\bigskip

\section{Introduction}

The inverse scattering transform method (ISM) is an
effective tool for studying nonlinear integrable equations
\cite{zah}. It allows one to construct large classes of
exact solutions and to investigate in detail the Cauchy
problem including the writing down explicit time
asymptotics. As for the mixed problem for the same
equations, here the success of the method is not so
impressive. For instance, the ISM applied to the initial
boundary value problem in its general formulation is not
sufficiently effective (see., for example, \cite{fi}),
except some special kinds of boundary conditions (see.
\cite{as,skl,tar,hab2}). The problem of searching boundary
conditions consistent with the integrability property of
the equation given as well as the problem of finding
proper modifications of analytical integration procedure
for the corresponding boundary value problems is
undoubtedly important.

Notice that in all known cases of initial boundary value problems studied up
by means of ISM the integration algorithm is based on reduction to the Cauchy
problem. One prolongs the initial value by means of the B\"acklund transform's
ordinary differential equations. This way is available if the equation is
invariant under a reflection type transformation $x\rightarrow-x$ only.

In our articles \cite{h}, \cite{ggh} a symmetry test was worked out to
verify whether the boundary value problem is consistent with integrability
or not. Utilizing this test in  \cite{ggh} and \cite{aggh} some examples of
boundary conditions were found for the Korteweg-de Vries type equations.
Evidently these equations don't admit any $x$ reflection. So to integrate
the corresponding initial boundary value problems the continuation algorithm
mentioned above is not suitable and in the present paper an alternative
approach is proposed. The essence of our approach consists of complementing
the conjugation contour of usual associated Riemann problem by adding new
branches. As an illustrative example we consider the following initial
boundary value problem for the Modified KdV equation
\begin{eqnarray}&& u_t=u_{xxx}+6\ve u_xu^2,\quad \ve=\pm1,
                                                                \label{kdv}\\
&&u\vert_{x=0}=0,\quad u_{x}\vert_{x=0}=0,
                                                                \label{x=0}\\
&&u\vert_{t=0}=u_0(x),\quad u_0(x)\vert_{x\ra+\infty}\ra0,
                                                                \label{t=0}
\end{eqnarray}
in the quadrant $x\geq0,$ $t\geq0.$ The consistency condition
$u_0(0)=u_{0x}(0)=0$ of the initial value with the boundary conditions is
supposed to be valid at the corner point $x=0,$ $t=0,$ and besides the
initial value $u_0(x)$ is required to vanish fast enough. The problem under
consideration is well posed. Notice that the correctness of mixed problems
like (\ref{kdv})-(\ref{t=0}) for linear and quasi-linear third order
equations is a well studied question (see, for instance, \cite{ton}). It is
worthwhile to say that the similar problem on the left semi-line $x<0$ is
not correctly posed -- in this case one should put only one boundary
condition at the point $x=0$. This fact implies immediately that the problem
(\ref{kdv})-(\ref{t=0}) can not be reduced to the Cauchy problem on a real
axis when the initial value $u_0(x)$ is chosen arbitrary.

Additionally we put rather severe requirement upon the initial data meaning
the absence of discrete eigenvalues of the auxiliary linear system of
equations (see conjectures i)-iv) in section 5 below) thus keeping out of
our attention an important question on soliton-like solutions to the
boundary value problem (\ref{kdv})-(\ref{x=0}). In this connection mention
the recent article \cite{ahs}, were an explicit description of
algebro-geometric solutions to the KdV equation is given satisfying
integrable boundary conditions.

Discuss briefly the content of the paper. In the second
section we consider a linear analogue of the problem
(\ref{kdv})-(\ref{t=0}) and construct its explicit
solution by the Wiener-Hopf method. In third one the Lax
representation is formulated for the initial boundary
value problem on a half-line. In fourth
-- the direct scattering problem is considered for the associated linear
system of equations. On a half-line solution of this problem is not unique in
contrast to the case of the whole line \cite{sha}. Some freedom arises even in
choosing the conjugation contour -- here it does not coincide with the
continuous spectra of the $x$-equation. In \S 5 it is proved that the
scattering matrix defines uniquely the coefficients of equation. The inverse
scattering problem is reduced to the Riemann problem posed on six rays
dividing the complex plane to six equal sectors. In \S 6 the time evolution of
scattering data is defined and it is shown that potential recovered in inverse
problem solves the initial boundary value problem (\ref{kdv})-(\ref{t=0}). In
the last seventh section the solvability of the Riemann problem on system of
rays is discussed.


\section{Solution of mixed problem in linear case}

This section is an auxiliary one, its aim is to explain with a simple
illustrative example the main parts of the algorithm proposed and in such a
way make easier the reading of the next sections.

Consider a linear version of the problem (\ref{kdv})-(\ref{t=0}):
\begin{eqnarray}&& u_t=u_{xxx},
                                                                \label{0kdv}\\
&&u\vert_{x=0}=0,\quad u_{x}\vert_{x=0}=0,
                                                                \label{0x=0}\\
&&u\vert_{t=0}=q(x),\quad q(x)\vert_{x\ra+\infty}\ra0.
                                                                \label{0t=0}
\end{eqnarray}
Apply to the equation (\ref{0kdv}) the Fourier transform instead of Laplace
one as it used to be. Set $v(\xi,t)=\int_0^{\infty}u(x,t)\exp (i\xi x) dx,$
then apparently the function $v(\xi,t)$ solves the equation
\bb
v_t(\xi,t)=u_{xx}(0,t)+i\xi^3v(\xi,t),                      \label{vt}
\ee
which contains an unknown function $u_{xx}(0,t)$. In order to
exclude the uncertainty arisen let us do a change of variables in
the equation putting $h(\xi,t)=v(\xi,t)-v(\o\xi,t),$ where
$\o=e^{{2i\pi\over3}}$ is a cubic root of the unity. The function
$h(\xi,t)$ satisfies much more simple equation $h_t=i\xi^3h$ which
is solved trivially. Now it is necessary to invert the change of
variables. To this end as well as in order to solve the scattering
problem below we need in the following geometric figures. Define on
the complex plane six congruent sectors $I_1,$ $I_2,...$ $I_6,$ by
dividing the complex plane by six rays $\ell_j,$ ended at the
origin: $\ell_j=\{\xi,\quad arg(\xi)={\pi\over 3}(j-1)\}$
enumerated in such a way that $I_j$ is a sector located between
rays $\ell_j$ и $\ell_{j+1}$ (see picture below)

\begin{figure}[h]                    \label{fig:pattern5}
\setlength{\unitlength}{0.05em}
\begin{picture}(200,410)(-350, -215)
 \put(-100, -150){\line( 2,3){200}} \put( 220,-10){$\ell_1$}
 \put(-240, -10 ){$\ell_4$}
 \put( 100, -150){\line(-2,3){200}} \put( 110,165){$\ell_2$}
 \put(-120, -175){$\ell_5$}
 \put(-200,    0){\line( 1,0){400}} \put(-120,165){$\ell_3$}
 \put( 110, -175){$\ell_6$}
 \put( 140,  70 ){$I_1$} \put( 140, -85 ){$I_6$}
 \put(-150,  70 ){$I_3$} \put(-150, -85 ){$I_4$}
 \put( -5 , 130 ){$I_2$} \put( -10, -140){$I_5$}
 \put(-140,-220 ){Picture. Conjugation contour.}
\end{picture}
\end{figure}
\noindent
Note that by construction the function $h(\xi,t)$ for $t=0$ is
analytical in sector $I_1$ and vanishes at infinity. These
properties are valid also for $t>0$, but violated for $t<0$. One
can check directly that the function searched is represented for
$t>0$ as a sum of three Cauchy type integrals
\bb
2\pi i v(\xi)=\int_{\ell_1}{h(s)ds\over s-\xi}
+\int_{\ell_4}{h(\o^2s)ds\over s-\xi} + \int_{\ell_5}{h(\o s)ds\over s-\xi}.
\label{v}
\ee
Here the integration is taken along rays $\ell_j$ in direction from
the origin to the infinity.  The condition $h(0)=0$ guarantees the
continuity of the function (\ref{v}) in the closed upper
half-plane.

The formula (\ref{v}) is simplified essentially when $\xi\in
I_1\cup I_2$. Really for such $\xi$ integration along the
ray $\ell_4$ can be replaced by integration along $\ell_3,$
and afterwards by means of the change of variables
$\nu=\xi^3$ one can reduce sum of three integrals in
(\ref{v}) to a integral along $\ell_1$:
$$
2\pi i v(\xi)=\int_{\ell_1}{h(\nu^{1/3})d\nu\over \nu-\xi^3}, \quad
\xi\in I_1\cup I_2.
$$
For $\xi\in I_3$ the function $v(\xi)$ is found from: $v(\xi)=
v(\o^2\xi) -h(\o^2\xi)$. Applying the inverse Fourier transform to
$v(\xi,t)$ one finds the solution searched $u(x,t)$ to the problem
(\ref{0kdv})-(\ref{0t=0}).


\section{Lax representation for the problem on a half-line}

Consider for convenience a system of equations in partial
derivatives containing equation (\ref{kdv}) as a
particular case
\bb u_t=u_{xxx}-6u_xuu^+,\quad u_t^+=u_{xxx}^+-6u^+_xuu^+.
                                                                     \label{ckdv}
\ee
Put in accordance with (\ref{x=0}) and (\ref{t=0}):
\bb
u|_{x=0}=u_x|_{x=0}=u^+|_{x=0}=u^+_x|_{x=0}=0                   \label{bc}
\ee
and
\bb
u(x,0)=u_0(x)\ra0, \quad u^+(x,0)=u_0^+(x)\ra0\quad \hbox{for}
\quad x\ra\infty.
                                                                        \label{ic}
\ee
All our constructions below are consistent with the involution
$u^+=-\ve u$ reducing the problem (\ref{ckdv})-(\ref{ic}) to the
problem (\ref{kdv})-(\ref{t=0}). The system of equations
(\ref{ckdv}) is a compatibility condition of the following over
determined system of equations
\begin{eqnarray}
&&y_x=(-i\xi\s_3+U)y, \label{yx}\\ && y_t=(4i\xi^3\s_3-4\xi^2U-\xi
V_1-V_0)y,
                                                                     \label{yt}
\end{eqnarray}
where
$$
V_1=\pmatrix{-2iuu^+&-2u_x\cr -2u^+_x&2iuu^+},
\quad
U=\pmatrix{0&iu\cr -iu^+&0},$$
$$ V_0=\pmatrix{-uu^+_x+u^+u_x&-iu_{xx}+2iu^2u^+\cr
iu_{xx}^+-2iuu^{+2}&uu^+_x-u^+u_x}.
$$
Along the line $x=0$ by means of the boundary conditions (\ref{bc})
equation (\ref{yt}) takes the form
\bb
y_t=(4i \xi^3 \s_3-U_{xx})y.
                                                                       \label{yt0}
\ee
Changing parameter $\nu=-4\xi^3$ in the last equation one gets
again Zakharov-Shabat equation of the form (\ref{yx}).

Thus three linear systems of equations (\ref{yx}), (\ref{yt}) and
(\ref{yt0}) provide the Lax representation for the initial boundary
value problem (\ref{ckdv}), (\ref{bc}) and (\ref{ic}).


\section{Direct scattering on a half-line}

Using the fact that the potential $U(x,0)$ vanishes
rapidly enough one can construct a matrix valued solution
$y(x,\xi)$ of the auxiliary system (\ref{yx}), by
prescribing for $x\ra +\infty$ the following
asymptotically trigonometric behavior
$$y_+(x,\xi)\rightarrow\exp(-i\xi x\s_3),$$
valid for any real $\xi.$ As it is known such a matrix
valued solution always exists and its columns
$y^{(1,2)}(x,\xi)$  define two vector valued functions
$\psi(x,\xi)=y^{1}(x,\xi)e^{i\xi x}$ and
$\phi(x,\xi)=y^{2}(x,\xi)e^{-i\xi x},$ admitting analytic
continuations with respect to the parameter $\xi$ from the
real axis onto the lower and, respectively, upper
half-planes. Moreover, the matrix valued function
$$(\psi(x,\xi),\,\phi(x,\xi))$$ tends to the unit matrix when ${\bf
R}\ni\xi\ra\infty.$ The matrix
$T(\xi)=(\psi(\xi),\,\phi(\xi))$, compound by columns
$\psi(\xi)=\psi(0,\xi)$ and $\phi(\xi)=\phi(0,\xi)$ is
called scattering matrix of the system (\ref{yx}) on a
half line.

Since to solve completely the direct scattering problem
means to find a pair of linearly independent piece-wise
analytic vector valued solutions to the system of
equations (\ref{yx}) which should be defined at all points
of the $\xi$-plane except may be a set of measure equal to
zero (see, for instance, \cite{sha}), hence we need in one
more solution. In our case the variable $x$ takes only
non-negative values, so we have a freedom in choosing the
second analytic solution. Utilize this freedom in the
following way. Compound matrix valued functions
$\Psi_{j}(\xi),$ analytic in sectors $I_j,$ with the help
of columns of the scattering matrix $T(\xi)$ by putting
dawn
\bb
\bba{lll}
\Psi_1(\xi)=(\psi(\o^2\xi),\,\phi(\xi)),\\
\Psi_2(\xi)=\Psi_3(\xi)=(\psi(\o\xi),\,\phi(\xi)),\\
\Psi_4(\xi)=\Psi_5(\xi)=(\psi(\xi),\,\phi(\o^2\xi)),\\
\Psi_6(\xi)=(\psi(\xi),\,\phi(\o\xi)).
\eea                                                                    \label{psi0}
\ee
Complete the discussion on the direct scattering problem defining the
solution $y(x,\xi)$ of the linear problem (\ref{yx}), by giving up its value
at the point $x=0$,
$$ y(0,\xi)=\Psi_j(\xi), \quad \xi\in I_j.$$
One can prove that the function $\Psi(x,\xi)=\Psi_j(x,\xi)$, given as
$$\Psi_j(x,\xi)=y(x,\xi)e^{i\xi x\s_3}, \quad \xi\in I_j,$$
is a piece-wise analytic function, defined on the whole complex $\xi$-plane.
In passing up from a sector $I_j$ to the sector $I_{j+1}$ the
$\psi$-function have a jupm:
\bb
\Psi_{j+1}(x,\xi)=\Psi_j(x,\xi)R_j(x,\xi).
                                                                     \label{rp}
\ee
The system of correlations (\ref{rp}) can be considered as
the Riemann problem on rays $\{\ell_j\}.$ Solutions to the
problem (\ref{rp}) satisfy some additional constraint
posed at the infinity $\Psi_j(x,\infty)=1$. Besides the
following asymptotic representation can be shown to be
valid
\bb
\Psi_j(x,\xi)=1+\sum_{k=1}^N \Psi_{jk}(x)\xi^{-k}+O(\xi^{-N-1}),
\quad I_j\ni\xi\ra\infty,                                       \label{nor}
\ee
if the coefficients $u_0(x)$, $u^+_0(x)$ of system of equations (\ref{yx}) are
smooth, rapidly decreasing functions. It can easily be checked that the
conjugation matrices depend on the variable $x$ in an explicit way
\bb
R_j(x,\xi)=e^{-i\xi x \s_3}R_j(0,\xi) e^{i\xi x \s_3}.          \label{rpx}
\ee


\section{Inverse scattering problem}

In the inverse scattering the problem is considered to
recover the potential $U(x)$ of the system (\ref{yx}) for
a scattering matrix given. Remind (see above) that the
scattering matrix is described by a pair of vector
functions $\psi(\xi),\,\phi(\xi),$ analytical in lower
and, respectively, upper half-planes. Coordinates of these
vectors are connected by equation
$\psi_1(\xi)\phi_2(\xi)-\psi_2(\xi)\phi_1(\xi)=1$ for
$\xi\in{\bf R}$. Because of the analytical properties of
entries this equation allows one to express functions
$\psi_1(\xi),$ $\phi_2(\xi)$ in terms of $\psi_2(\xi)$ and
$\phi_1(\xi)$ by well known formulae (see \cite{zah}), if
zeros of functions $\psi_1(\xi),$ $\phi_2(\xi),$ located
in domains $Im\xi<0$ and $Im\xi>0,$ respectively, are
listed a priori, and $1+\psi_2(\xi)\phi_1(\xi)\neq0$ for
$Im
\xi=0$.

Note that the scattering matrix of the half-line problem defines
uniquely the potential of system of equations (\ref{yx}) in
contrast to the case of whole line. Let us formulate a precise
statement.

{\bf Proposition 1.} Let $T(\xi)$ and $\ti T(\xi)$ are scattering
matrices, corresponding the potentials $U(x)$ и $\ti U(x).$ Then
the equation $T(\xi)=\ti T(\xi)$ implies the equation $U(x)=\ti
U(x).$

{\bf Scheme of proof.} Suppose that $T(\xi)=\ti T(\xi)$. It means
that the Riemann problem (\ref{rp}) has two different solutions
$\Psi_{j+1}(x,\xi)=\Psi_j(x,\xi)R_j(x,\xi)$ and
$\ti\Psi_{j+1}(x,\xi)=\ti\Psi_j(x,\xi)R_j(x,\xi).$ Let $f(x,\xi)$
be a ratio of these two solutions
$$ f(x,\xi)=\ti\Psi_j(x,\xi)\Psi_j^{-1}(x,\xi)=\ti\Psi_{j+1}(x,\xi)
\Psi_{j+1}^{-1}(x,\xi).$$
Evidently function $f(x,\xi)$ is bounded for large values
of $\xi,$ satisfies a linear equation with respect to $x$:
$f_x=i\xi[f,\s_3]+uf\s_2-\ti u\s_2f,$ besides
$f(0,\xi)=1.$ Consequently, it is analytical on whole
complex plane and according to the Liouville theorem it
doesn't depend on $\xi$. It follows from the linear
equation on $f$ that $[f,\s_3]=0.$ Thence, $f_x=0$ and
therefore $f(x,\xi)\equiv1,$ that means that $U(x)=\ti
U(x).$

Exclude from our consideration some degenerate cases in
the inverse scattering problem by imposing the following
additional requirements:
\begin{eqnarray}
i)&& 1+\psi_2(\xi)\phi_1(\xi)\neq0\quad \hbox{for}
\quad \xi\in {\bf R},\nonumber\\
ii)&& \log(1+\psi_2(\xi)\phi_1(\xi))
 |^{\xi=+\infty}_{\xi=-\infty}=0,\nonumber\\
iii)&& \langle \phi(\xi),\psi(\o^2\xi) \rangle\neq0\quad
\hbox{for} \quad \xi\in \bar I_1\cup\bar I_2,\nonumber\\
vi)&&\langle \phi(\xi),\psi(\o\xi) \rangle\neq0\quad
\hbox{for} \quad \xi\in \bar I_2\cup\bar I_3,\nonumber
\end{eqnarray}
where the skew-scalar product $\langle X,Y\rangle$ between two-dimensional
vectors $X$ and $Y$ is understood as the determinant of matrix: $\langle
X,Y\rangle =\det(X,Y)$ and the bar over a letter denotes closure of a
multitude.

{\bf Proposition 2.} Assume we are given a pair of functions
$\phi_1(\xi),$ $\psi_2(\xi)$ such that
$$ \phi_1(\xi)=\int^{+\infty}_0\exp (i\xi x )f_1(x)dx,\quad
\psi_2(\xi)=\int^{+\infty}_0\exp (-i\xi x )f_1(x)dx,$$
where $f_{1,2}(x)$ vanish fast enough together with all
derivatives. Recover the matrix $T(\xi)=(\psi(\xi),\,\phi(\xi))$,
choosing the first coordinate of the vector $\psi$ and the second
coordinate of the vector $\phi$ as Cauchy type integrals
$$\psi_1(\xi)=\exp (-{1\over 2\pi i}\int ^{+\infty}_{-\infty}
{\log(1+\psi_2(s)\phi_1(s))\over s-\xi}ds),\quad Im\xi<0,$$
$$\phi_2(\xi)=\exp ({1\over 2\pi i}\int ^{+\infty}_{-\infty}
{\log(1+\psi_2(s)\phi_1(s))\over s-\xi}ds),\quad Im\xi>0.$$ Let the all
conditions i)-vi) are valid. Then matrix $T(\xi)$ is a scattering matrix of
the problem on a half-line if and only if for all $x\geq0$ a non-degenerate
solution ($\det
\Psi_j\neq0$ for all $\xi\in \bar I_j$) exists to the Riemann problem
(\ref{rp}) with the conjugation matrices defined as
\begin{eqnarray}
&&R_1(0,\xi)={1\over A_{22}(\xi)}\pmatrix{-A_{11}(\o^2\xi) &0\cr
-A_{21}(\xi)&A_{22}(\xi)},\nonumber\\
                        \nonumber\\
&&R_3(0,\xi)={1\over A_{11}(\o^2\xi)}\pmatrix{1 & A_{12}(\o^2\xi)\cr
A_{21}(\o^2\xi)&-A_{22}(\o^2\xi)},\nonumber\\
                                \nonumber\\
&&R_5(0,\xi)={1\over A_{11}(\o\xi)}\pmatrix{A_{11}(\o\xi) & A_{12}(\o\xi)
\cr 0&-A_{22}(\o\xi)},\nonumber\\
                                        \nonumber\\
&&R_6(0,\xi)={1\over A_{22}(\o\xi)}\pmatrix{-A_{11}(\xi) &-A_{12}(\xi)\cr
-A_{21}(\o\xi)&-1} ,  \nonumber\\
                        \nonumber\\
&&R_2=R_4=1,          \nonumber
\end{eqnarray}
where
\begin{eqnarray}
&A_{11}(\xi)=\langle \psi(\o^2\xi),\,\phi(\o\xi)\rangle, \quad
&A_{12}(\xi)=\langle
\phi(\xi),\,\phi(\o\xi)\rangle, \nonumber\\ &A_{21}(\xi)=\langle
\psi(\o^2\xi),\,\psi(\o\xi)\rangle, \quad &A_{22}(\xi)=\langle
\phi(\xi),\,\psi(\o^2\xi)\rangle.\nonumber
\end{eqnarray}

{\bf Scheme of proof.} The necessity of the conditions
(\ref{rp}), (\ref{rpx}) was proven in the previous
section. Prove now the sufficiency. Define the conjugation
matrices $R_j$ for the scattering matrix
$T(\xi)=(\psi(\xi),\,\phi(\xi))$ given as it is pointed
out in the proposition. Suppose that for all $x\geq0$ the
Riemann problem (\ref{rp}), (\ref{rpx}) is uniquely
solvable. Find the potential $U(x)$ of the system
(\ref{yx}). To this end differentiate equation (\ref{rp})
with respect to $x$ and then transform it to the form
$$
\Psi_{j+1,x}\Psi^{-1}_{j+1}-i\xi\Psi_{j+1}\s_3\Psi^{-1}_{j+1}=
\Psi_{j,x}\Psi^{-1}_{j}-i\xi\Psi_{j}\s_3\Psi^{-1}_{j}=g.
$$
Applying now the Liouville theorem and asymptotic representation (\ref{nor})
it is easy to get that $g=-i\xi\s_3+U,$ i.e.
\bb
\Psi_{j,x}=-i\xi[\s_3,\Psi_j]+U\Psi_j.                               \label{psix}
\ee

{\bf Remark to the proposition 2.} The functions $A_{ij}$
and conjugation matrices $R_j$ admit some analytical
properties. For instance, $A_{11}$ is analytic in the
domain $I_1\cup I_6$, $A_{12}$ -- in the sector $I_1$,
$A_{21}$ -- in the sector $I_2$ and, lastly, $A_{22}$ is
analytic in $I_1\cup I_2.$ Conjugation matrices $R_1$ and
$R_5$ are analytic, respectively, in $I_2$ and $I_5$. The
matrices $R_3$ and $R_6$ are defined, generally speaking,
only on corresponding contours $\ell_4$ and $\ell_1$.


\section{Time dynamics and boundary conditions}

The scattering matrix evaluates in time in a very
complicated way -- it is the main difficulty in
investigating the initial boundary value problem. One can
get some impressions about the problem by looking at the
equation (\ref{vt}) on the function $v(\xi,t)$ in the
linear case. On the other hand side the conjugation
matrices $R_j$, which are some rational functions of the
elements of the scattering matrix, depend on $t$
explicitly. It is more convenient to postulate time
dynamics in the inverse scattering problem than to find it
up in the direct problem. Let the conjugation matrices
$R_j(x,\xi,t)$ of the Riemann problem (\ref{rp}),
(\ref{rpx}) evaluate in time by means of the following
equations $R_{j,t}=4i\xi^3[\s_3,R_j]$, so that
\bb
R_j(x,\xi,t)=e^{4i\xi^3t\s_3}R_j(x,\xi) e^{-4i\xi^3t\s_3}.
                                                                        \label{rpt}
\ee
Differentiate the equation (\ref{rp}) with respect to $t$,
and then replacing by means of (\ref{rpt}) write it down
in the form
$$
\Psi_{j+1,t}\Psi^{-1}_{j+1}+4i\xi^3\Psi_{j+1}\s_3\Psi^{-1}_{j+1}=
\Psi_{j,t}\Psi^{-1}_{j}+4i\xi^3\Psi_{j}\s_3\Psi^{-1}_{j}=h.
$$
Applying the Liouville theorem to the function $h$ and
afterwards substituting the asymptotic representation
(\ref{nor}) one can find that $h$ takes the form
$4i\xi^3\s_3-4\xi^2U-\xi V_1-V_0$ (see formula
(\ref{yt})). Consequently, $\Psi_j$ solves the equation
\bb
\Psi_{j,t}=4i\xi^3[\s_3,\Psi_j]-(4\xi^2U+\xi
V_1+V_0)\Psi_j.
                                                        \label{psit}
\ee

As it follows from equations (\ref{psix}), (\ref{psit})
the function
$$
y(x,\xi,t)=\Psi_j(x,\xi,t)e^{(4\xi^3t-\xi x)i\s_3}
$$
is a solution to the over-determined system of equations
(\ref{yx}), (\ref{yt}), hence the potential $U(x,t)$,
found by solving the Riemann problem contains as a
component a solution to the system of nonlinear equations
(\ref{ckdv}), moreover, the initial conditions are
automatically valid (\ref{ic}). Check the validity of the
boundary conditions (\ref{bc}). To this end it is enough
to show that at the point $x=0$ the system of equations
(\ref{yt}) is invariant under rotation $\xi\ra
\o\xi$, i.e. not only the functions $\Psi_j(\xi,t)$, but also functions
$\Phi_j(\xi,t)=
\Psi_{j+2}(\o\xi,t)$ are solutions to the system (\ref{psit}).
For values of the index $j$ which are greater than six the
function $\Psi_j$ is defined periodically on $j$ with
period 6. By construction for any $j$ the function
$\Phi_j(\xi,t)$ is analytic in the sector $I_j$, as well
as, the function $m_j(\xi,t)=\Psi_j^{-1}(\xi,t)
\Phi_j(\xi,t)$ (requirements i)-iv), are supposed to still valid, therefore
$\det \Psi_j\neq0$ for $\xi\in \bar I_j$). As a result of
this definition one gets a new Riemann problem similar
(\ref{rp}),
\bb
\Phi_{j+1}(\xi)=\Phi_j(\xi)r_j(\xi),\quad r_j(\xi)=R_j(\o\xi),
                                                             \label{rpf}
\ee
with conjugation matrices
$r_j(\xi)=m_j(\xi)^{-1}R_j(\xi)m_{j+1}(\xi)$. When $t=0$
it can directly be shown that matrices $m_1,\, m_3,\, m_5$
are upper-triangular, and matrices $m_2,\, m_4,\, m_6$ are
lower-triangular. By beans of this triangular structure
the functions
$$
\ti m_j(\xi,t)=e^{4i\xi^3\s_3t}m_j(\xi,0)e^{-4i\xi^3\s_3t}
$$
are also analytic in the corresponding sectors $I_j$. But
it means that functions $\ti\Phi_j
(\xi,t)=\Psi_j(\xi,t)\ti m_j(\xi,t)$ are also solutions to
the same Riemann problem (\ref{rpf}), which is uniquely
solvable, hence the functions $\ti\Phi_j(\xi,t)$ and
$\Phi_j(\xi,t)$ should coincide, consequently, $\ti
m\equiv m$. Therefore the product
$(\Phi_je^{4i\xi^3\s_3t})\times
(\Psi_je^{4i\xi^3\s_3t})^{-1}$ is equal to $m(\xi,0)$, or
in other words, it doesn't depend on $t$, i.e. functions
$y(\xi,t)=\Psi_je^{4i\xi^3\s_3t}$ and $\ti
y(\xi,t)=\Phi_je^{4i\xi^3\s_3t}$ solve one and the same
equation of the form (\ref{yt}). By construction one has
$\ti y(\xi,t)=y(\o\xi,t)$, consequently, at the point
$x=0$ equation (\ref{yt}) is not changed under rotation
$\xi\ra\o\xi$, but then it is of the form (\ref{yt0}). It
is equivalent to the validity of the boundary condition
(\ref{bc}).

The conjugation matrices $R_1$ and $R_5$ as triangular
ones (see proposition 2 and remark to it) remain
analytical in the corresponding sectors $I_2$ and $I_5$
for all $t>0$.


\section{On solvability of the Riemann problem on a system of rays}

The Riemann problem on analytical conjugation on a system
of rays which the inverse scattering is reduced to is not
sufficiently well studied in spite of some close
formulations has been met earlier in the connection with
the Painleve equations \cite{jm}. But in our case the
Riemann problem has inner symmetries which allow one to
reduce it to the well known version of the problem posed
on the real axis by transformations different for $x=0$
and for $x>0$.

Let us begin with the case $x=0$, $t\geq 0$. Note first
that functions $\Psi_1(\xi,t)$ and $\Psi_6(\xi,t)$ are
analytic not only in the corresponding sectors $I_1$ and
$I_6$ but also in larger domaims such as $I_1\cup I_2$ and
$I_5\cup I_6$, respectively, moreover they are related to
each other by an involution
$\Psi_6(\xi,t)=\Psi_1(\o\xi,t)$. Define one more function
by setting $\Psi_{ex}(\xi,t)=\Psi_1(\o^2\xi,t)$. These
three functions constitute a solution to the following
Riemann problem
\bb
\Psi_1=\Psi_6M_6,\quad \Psi_6=\Psi_{ex}M_{ex},\quad
\Psi_{ex}=\Psi_1M_1\quad                              \label{rpex}
\ee
on a system of three rays $\ell_1$, $\ell_3$, $\ell_5$ and
in addition satisfy one and the same system of equations
of the form (\ref{psit}). Hence the conjugation matrices
$M_1$, $M_6$, $M_{ex}$ evaluate in time $t$ as follows
\bb
M(\xi,t)=e^{4i\xi^3\s_3t}M(\xi,0)e^{-4i\xi^3\s_3t}.      \label{m}
\ee
It can be checked that the Riemann problems (\ref{rp}),
(\ref{rpt}) and (\ref{rpex}), (\ref{m}) are equivalent:
solution to one of them can easily be transformed into a
solution to the other one by a simple explicit way. Let us
given, for instance, a solution to the problem
(\ref{rpex}). Then functions $\Psi_2,$ $\Psi_3,$ $\Psi_4,$
$\Psi_5$ which are the solution to the problem (\ref{rp})
are expressed as
$$
\Psi_2=\Psi_1 R_1,\quad \Psi_3=\Psi_{ex}N_3,\quad\Psi_4=\Psi_{ex}N_4,
\quad\Psi_5=\Psi_6 R_5^{-1},\quad
$$
where $N_3$ and $N_4$ are upper- and lower-triangular
matrices. It is easy to check when $t=0$ that $N_3$ and
$N_4$ are analytic in $I_3$ and $I_4$ and therefore, by
means of the equations
$N_{3,4}(\xi,t)=e^{4i\xi^3\s_3t}N_{3,4}(\xi,0)e^{-4i\xi^3\s_3t}$
and triangular structure they preserve their analytical
properties for $t>0$ also.

Change variables $\xi=\nu^{1/3}$ in the problem
(\ref{rpex}). Evidently if $\xi$ varies in the complex plane
then $\nu$ varies in three sheets Riemannien surface however
in all these sheets the function defined as
$\Psi(\nu,t)=\Psi_1(\xi,t)$ takes one and the same values,
i.e. it has no branch points, but fulfills a jump when
crosses the positive half-line $\nu>0$. Under the change of
variables the Riemann problem on three rays is brought to
the problem on real axis
\bb
\Phi(\nu,t)\Psi(\nu,t)=C(\nu,t),\quad \nu\in {\bf R},       \label{rpex2}
\ee
where $\Phi(\nu,t)=-A_{22}(\o\xi)\Psi_6^{-1}(\xi,t)$ and the
conjugation matrix is chosen as $C(\nu,t)=-A_{22}(\o\xi)$
for $\nu\leq0$, $(\xi\in\ell_2)$ и
$C(\nu,t)=-A_{22}(\o\xi)R_6(0,\xi,t)$ for $\nu>0$
$(\xi\in\ell_2)$. Thus in order to solve the problem
(\ref{rp}) it is enough to solve the problem (\ref{rpex2}).
Now it becomes clear that the problem (\ref{rp}) is solvable
if functions $\psi_2(\xi)$, $\phi_1(\xi)$ are close to zero.
This follows from the famous theorem about factorisation on
a neighbourhood of the unity (see, for instance,
\cite{pre}). The positivity of the conjugation matrix also
provides the unique solvability of the problem
(\ref{rpex2}). Assume that the reduction constraint
$u^+(x,t)=-\ve u(x,t)$, $\ve=\pm1$ is imposed on the initial
boundary value problem (\ref{ckdv}), (\ref{bc}), (\ref{ic}).
Then columns $\psi$ and $\phi$ of the scattering matrix $T$
satisfy the involution $\phi_2(\xi)=\psi_1^*(\xi^*)$
$\phi_1(\xi)=-\ve\psi_2^*(\xi^*)$, where the star over a
letter denotes the complex conjugation. In this case the
conjugation matrix of the problem (\ref{rpex2}) owing to the
constraints i)-iv) from \S5 is positively defined. Actually,
the reduction condition gives
$A_{22}^*(\o\xi)=A_{22}(\o\xi)$ for $\xi\in\ell_2$.
Moreover, $A_{22}(0)=A_{22}(\infty)=-1$, $A_{22}(\xi)\neq0$,
hence, $A_{22}(\xi)<0$ along the ray $\ell_2$ such that
$C(\nu,t)>0$ for $\nu\leq0$. Compute now the principal
minors of the matrix $C(\nu,t)$:
$\det_1C(\nu,t)=A_{11}(\xi),$ $\det_2C(\nu,t)=\det
C(\nu,t)=|A_{22}(\xi)|^2$ for $\nu\geq0$. By means of the
constraint mentioned both are positive. Hence $C(\nu,t)$ is
positively defined and, besides, it is continuous in the
H\"older sense that provides the unique solvability of the
problem (\ref{rpex2}).

Consider now the case $0<x<\infty$, $t=const>0$, supposing
that the Riemann problem (\ref{rp}), (\ref{nor}),
(\ref{rpx}), (\ref{rpt}) is already solved for $x=0$,
$t>0$. It means really that the scattering matrix
$T(\xi,t)= (\psi(\xi,t),
\phi(\xi,t))$ of the problem on the half-line $x\geq0$ is
known for all $t>0$ and we come to the problem of
recovering of the potential of the system (\ref{yx}) for a
given scattering matrix. The only solution of this problem
(on the uniqueness see above proposition 1) can be found
in two ways: by means of the Riemann problem (\ref{rp}),
(\ref{nor}), (\ref{rpx}) on a system of rays and by the
canonical method based on the same problem on the real
axis \cite{sha}:
\begin{eqnarray}
&&\hat\phi(x,\xi,t)=\hat\psi(x,\xi,t)
\pmatrix{1&\phi_1(\xi,t)\exp (-2i\xi x)
\cr -\psi_2(\xi,t)\exp (2i\xi x)&1},\nonumber\\
&&\hat\phi(x,\infty,t)=1.
                                                                \label{crp}
\end{eqnarray}
For $x\geq0$ and for $t\geq0$ the Riemann problems
mentioned are equivalent unless the conditions i)-iv) are
violated. Solutions to these two problems are related to
each other by equations
$$
\hat\phi(x,\xi,t)=\psi_1(x,\xi,t)m_1(x,\xi,t), \quad
\hat\phi(x,\xi,t)=\psi_3(x,\xi,t)m_3(x,\xi,t), \quad
$$
$$
\hat\psi(x,\xi,t)=\psi_5(x,\xi,t)m_5(x,\xi,t), \quad
\hat\psi(x,\xi,t)=\psi_6(x,\xi,t)m_6(x,\xi,t), \quad
$$
where $m_j(x,\xi,t)$ are triangular matrix-valued
piece-wise analytic functions. As an example we give an
explicit representation of the matrix $m_3(x,\xi,t)$:
$$
m_3(x,\xi,t)=\Psi_3^{-1}\hat\phi={1\over
A_{11}(\o^2\xi)}\pmatrix{\phi_2(\xi,t)&0
\cr-\psi_2(\o\xi,t)e^{2ix\xi}&A_{11}(\o^2\xi)}.
$$
It is well known (see, for instance, \cite{sha}) that
under reduction $u^+(x,t)=-\ve u(x,t)$, $\ve=\pm1$ and the
conditions i)-ii) the Riemann problem (\ref{crp}) is
uniquely solvable.

It is worth to add that in the case $\ve=-1$ the
conditions i)-iv) don't impose any constraint on the
initial value of the initial boundary value problem
(\ref{kdv})-(\ref{t=0}), they are always valid.


\begin{thebibliography}{99}

\bibitem{zah} V.E.Zakharov, S.V.Manakov, S.P.Novikov, L.P.Pitaevskij.
        Soliton theory. Inverse scattering method. М.: Nauka, 1980 (in Russian).

\bibitem{fi} A.S.Fokas, A.R.Its. TMF. 1992. V.92. no 3, P.386-403.

\bibitem{as} M.J.Ablowitz, H.J.Segur. Math. Phys. 1975. V. 16. P. 1054.

\bibitem{skl} E.K.Sklyanin. Funct. anal. i priloj. 1987. V. 21. no 2.
        P. 86-87 (in Russian).

\bibitem{tar} V.O.Tarasov. Inverse Problems. 1991. V. 7. P. 435.

\bibitem{hab2} I.T.Habibullin. Teoret. matem. fizika, 1991. V. 86.
        no 1. P. 43-51 (in Russian).

\bibitem{h} I.T.Habibullin. Phys. Letts. A. 1993. V. 178. P. 369.

\bibitem{ggh} B.G\"urel, M.G\"urses, I.Habibullin.
        J. Math. Phys. 1995. V. 36. no 12. P. 6809.

\bibitem{aggh} V.Adler, B.G\"urel, M.G\"urses, I.Habibullin.
        J. Phys. A: Math. Gen. 1997. V.30, P. 3505-3513.

\bibitem{ton} M.D.Ramazanov. Matem. sbornik. 1964. V. 64. no 2. P. 234.
         An Ton Bui. J. Diff. Eq. 1977. V. 25. no 3. P. 288. A.V.Faminskij.
        Trudy MMO. 1988. V. 51. P. 54 (in Russian).

\bibitem{ahs} V.E.Adler, I.T.Habibullin, A.B.Shabat. TMF. 1997. V.110.
        no 1. P. 98-113.

\bibitem{sha} A.B.Shabat. Differen. Uravnen. 1979. V. 15. no 10.
        P. 1824-1835.

\bibitem{pre} Z.Pr\"ossdorf. Einige Klassen singul\"arer Gleichungen.
        Akademie-Verlag-Berlin, 1974.

\bibitem{jm} M.Jimbo, T.Miwa. Physica D. 1981. V.2. P.407-448.

\end{thebibliography}
\end{document}